%% file: stability_techreport.tex
\documentclass[
11pt,
DIV=18
]{scrartcl}
\pdfoutput=1 

\usepackage{amsfonts}
\usepackage{amsmath}   
\usepackage{amssymb}
\usepackage{color}
\usepackage{graphicx} 
\usepackage[]{algorithm}
\usepackage{algpseudocode}
\usepackage[colorlinks=false, pdfborder={0 0 0}, breaklinks=true]{hyperref}
\usepackage[utf8]{inputenc}
\usepackage{icomma}
\usepackage{units}
\usepackage[T1]{fontenc}
\usepackage{listings}

\DeclareUnicodeCharacter{00A0}{~}

\begin{document} 
\title{A parallel and matrix free framework for global stability analysis of compressible flows}
\author{O. Henze, M. Lemke, J. Sesterhenn \\ \small{Institut f\"ur Str\"omungsmechanik und Technische Akustik, Technische Universit\"at Berlin}} 
\date{}
\graphicspath{{figures/}}

\maketitle

\begin{abstract}
An numerical iterative  framework for global modal stability analysis of compressible flows using a parallel environment is presented. The framework uses a matrix-free implementation to allow computations of large scale problems. Various methods are tested with regard to convergence acceleration of the framework. The methods consist of a spectral Cayley transformation used to select desired Eigenvalues from a large spectrum, an improved linear solver and a parallel block-Jacobi preconditioning scheme. \\

\noindent stability analysis $\cdot$ compressible flow $\cdot$ matrix free methods
\end{abstract}

\input{01_introduction}

\input{02_numerical_framework}

\input{03_stability_analysis}
\input{04_results}

\bibliographystyle{alphaabbr}
\bibliography{stability_techreport}

\end{document}

%% file: 01_introduction.tex
\section{Introduction}

In fluid dynamics, it is crucial for many technical applications to obtain stability information of the flow. While the stability of certain flows like the inviscid hyperbolic-tangent velocity profile is well studied \cite{Blumen1970, Michalke1964}, the global stability analysis of general flows is still a daunting task.
Traditionally a stability analysis by e.g. ARPACK \cite{Lehoucq1998} is performed by obtaining a linearized matrix representation of the discretized operator and using it to perform the analysis. 

This kind of computation is usually limited by the available memory and computational resources. But certain physical phenomena require a highly resolved discretization and high order discretization schemes, for example the computation of acoustic modes generated by a compressible flow. This leads to an immense memory requirement during a global stability analysis, since it requires information about the describing operator in the whole computational domain. This encourages the use of a matrix free method to perform the stability analysis, which lessens the required memory significantly.

This article presents a framework which allows a matrix free computation of a global linear temporal stability analysis of compressible flows using a high order spatial discretization of the compressible Navier-Stokes equations. Compared to previous applications of this method, for example by \cite{Mack2009}, the numerical stability analysis can be performed in a parallel manner.  

We focus on the the temporal linear stability analysis of compressible flows. Being able to compute the stability analysis without the need of an operator in matrix formulation leads to a considerable reduction in required memory during a computation. In turn, certain challenges which arise because of the matrix free method have to be considered.

The presented framework employs iterative Krylov methods in order to obtain the desired Eigeninformation. Krylov methods are a natural choice for this kind of problem since they can be parallelized well and can operate without explicit knowledge of the operator. The base method of choice is the implicitly restarted Arnoldi method (IRAM) \cite{Sorensen2002}. It is combined with a spectral Cayley transformation to facilitate in the selection of the correct Eigenvalues and to provide a speedup of the computation \cite{Meerbergen1994}. The use of this transform introduces a new linear system which has to be solved. An improvement of a common iterative method (GMRES) is tested as well as a preconditioning scheme in order to improve convergence.

Validation of the method is performed on a compressible mixing layer setup using a tangent hyperbolic profile as investigated by Blumen \cite{Blumen1970}. The growth rate of the most unstable mode of this profile is computed using the proposed framework and compared to literature. 

This article is structured  as follows: Section $2$ of this paper will discuss the physical model used in the stability analysis. Section $3$ will explain in detail the concepts used in the stability analysis framework and the parallelization of it. To close, a model problem and results based on it are presented and evaluated in sections $4$ and $5$.

%% file: 02_numerical_framework.tex
\section{Governing equations}

For the problems considered herein, the compressible Navier-Stokes \eqref{eq:navier_stokes} equations are used as the governing equations. They are given in a characteristic type formulation \cite{Sesterhenn2000} and are used in a matrix free fashion. Note that the use of this special formulation is not required for the framework, other formulations can be applied as well. The formulation introduces the primitive variables $p$ as pressure, $u_{i}$ as the velocity vector and $s$ as entropy. Additionally, $\tau_{ij}$ describes the stress tensor, $T$ the temperature, $\Phi$ the dissipation and $\rho$ the density, with $\mu$ and $\mu_{v}$ describing the viscosity and bulk viscosity of the fluid, respectively.
\begin{align}
  \begin{aligned}
    \frac{\partial p }{\partial t} + \varrho c^2 \frac{\partial u_{i} }{\partial x_{i}} &= \frac{p}{c_v} \left(\frac{\partial s}{\partial t} + u_{i} 
    \frac{\partial s}{\partial x_{i}} \right)  \\
    \frac{\partial u_{i} }{\partial t} + u_{j} \frac{\partial u_{i} }{\partial 
    x_{j}} &= - \frac{1}{\varrho} \frac{\partial p}{\partial x_{i}} + \frac{1}
    {\varrho} \frac{\partial \tau_{ij}}{\partial x_{j}}  \\
    \frac{\partial s }{\partial t} + u_{j} \frac{\partial s }{\partial x_{j}} 
    &= -\frac{1}{\varrho T} \left( \frac{\partial }{\partial x_{j}} \left( 
    \lambda \frac{\partial T}{\partial x_{j}} \right) + \Phi \right)  \\
    \Phi & = \tau_{ij} s_{ij} \\
    \tau_{ij} & = 2 \mu \left( s_{ij} - \frac{1}{3} s_{kk} \delta_{ij} \right) + \mu_v s_{kk} \delta_{ij} \\
    s_{ij} & = \frac{1}{2} \left( \frac{\partial u_i }{\partial x_j} +\frac{\partial u_j }{\partial x_i} \right)
  \end{aligned} \label{eq:navier_stokes}
\end{align}
This set of equations is closed with Sutherland's law and the ideal gas equation and uses the summation convention with $i = 1,2,3$.

Going forward, the short hand operator $\mathcal{N}(q)$ will be used to refer to the Navier Stokes equations, with $q$ containing all primitive flow variables $q = \left[p, u_{i} , s \right]$.

Within the used scheme, the wave propagation terms are discretized with a sixth order finite difference scheme using compact finite difference schemes by \cite{Lele1992} on a staggered grid, which allows an accurate computation required to extract the stability information. The dissipation terms were discretized using a similar scheme of fifth order.  \par

%% file: 03_stability_analysis.tex
\section{Numerical stability analysis}
\subsection*{Ansatz and overview}

In order to perform a global stability analysis, we need to formulate an appropriate ansatz which allows the computation of the temporal stability properties of the flow. 
\begin{align}
  \dot{q} = {J}(q_0) q \label{eq:model_stabiity}
\end{align}
Therein ${J}(q_0)$ describes the temporal evolution of the operator $\mathcal{N}_{\mathrm{lin}} = J(q_0)$, linearized around a base state $q_o$.
To this end we investigate the temporal stability.
The goal of the stability analysis is to obtain the Eigenvalues $\lambda$ of the Operator. For this, an ansatz fitting to the problem has to be chosen. The form of the ansatz depends on the structure of the investigated flow, especially on its degree of homogeneity. For a general three dimensional flow without any degree of spatial homogeneity, the general approach \eqref{eq:disturbance_mack} can be used for the stability analysis, as given by \cite{Mack2009}.
\begin{align}
  \tilde{q}(x,y,z,t)  =  \phi(x,y,z) \exp(- i \alpha c t ) \label{eq:disturbance_mack}
\end{align}
Herein, $\alpha$ denotes the wave number of the temporal disturbance, $c$ represents the wave propagation speed,  while $\phi(x,y,z)$ describes the Eigenvector corresponding to $\alpha$. Combining the model equation \eqref{eq:model_stabiity} with the general approach \eqref{eq:disturbance_mack} results in an Eigenvalue problem \eqref{eq:stability} which has to be solved.
\begin{align}
 J(q_{0}) \phi(x,y) = \lambda \phi(x,y) \label{eq:stability} 
\end{align}

An Eigenvalue problem of this type can be solved directly if the Jacobian matrix $J(q_{0})$ is known. Since the memory requirements for storing this matrix are too high for global stability analysis problems, we have to employ matrix free methods to solve the it.

\subsection*{Matrix free framework}
In order to compute the Eigenvalues needed for the stability analysis, the Jacobian of the operator solving the defining equations \eqref{eq:navier_stokes} is required. Since the defining equations are nonlinear and not available in a matrix formulation, an approximation of the Jacobian has to be computed. Therefore a Frechèt-type derivative \eqref{eq:frechet} is employed for the linearization of the nonlinear operator during the stability analysis.
\begin{align}
 J(q_0) q = \frac{\mathcal{N}(q_0 + \epsilon q) - \mathcal{N}(q_0)}{\epsilon} \approx  \mathcal{N}_{Lin} q \label{eq:frechet}
\end{align}

The correct choice of the parameter $\epsilon$ required for this derivative is critical to its accuracy. We used the findings on the influence of $\epsilon$ given by \cite{Schulze2009} and set it accordingly to $\epsilon = \|q\| \sqrt{\epsilon_{\mathrm{m}}}$, with $\epsilon_{\mathrm{m}}$ representing the machine accuracy of the system.
The use of this linearization requires the evaluation of the defining equations $\mathcal{N}(q)$ every time the Jacobian is required during the computation. This application is usually one of the most numerically expensive operations and one should therefore try to minimize the number of required calls.

The use of a matrix free method to obtain information about $J$ without explicitly storing it leads naturally to the use of iterative Krylov subspace methods, which do not require the Jacobian in explicit form but rather just matrix vector products of the type  $J(q_{0}) \phi(x,y)$. These matrix vector products can be computed in a matrix free fashion by using the previously described Frechèt derivative \eqref{eq:frechet}. 

\subsection*{Krylov methods}
Since the framework is designed to work in a matrix free environment, a method which can operate in such an environment is required for the stability analysis. Krylov methods suite this requirement well and will therefore be used as method of choice.
The idea of Krylov methods is to iteratively create a sequence of subspaces $\mathcal{K}_{m}$ which meets \eqref{eq:krylov}.
\begin{align}
 \mathcal{K}_{1}(\mathcal{N}_{\mathrm{lin}},r_0) \subset \mathcal{K}_{2}(\mathcal{N}_{\mathrm{lin}},r_0)\subset ... \subset \mathcal{K}_{m}(\mathcal{N}_{\mathrm{lin}},r_0) \label{eq:krylov}
\end{align}
As stated before, $\mathcal{N}_{\mathrm{lin}} = J(q_o)$ is a (linearized) matrix representation of the Jacobian of $\mathcal{N}(q)$. Note that due to the iterative nature of Krylov methods only the product given by the Frechèt derivative \eqref{eq:frechet} is required.

\begin{figure}
  \centering
  \includegraphics[width=0.7\textwidth]{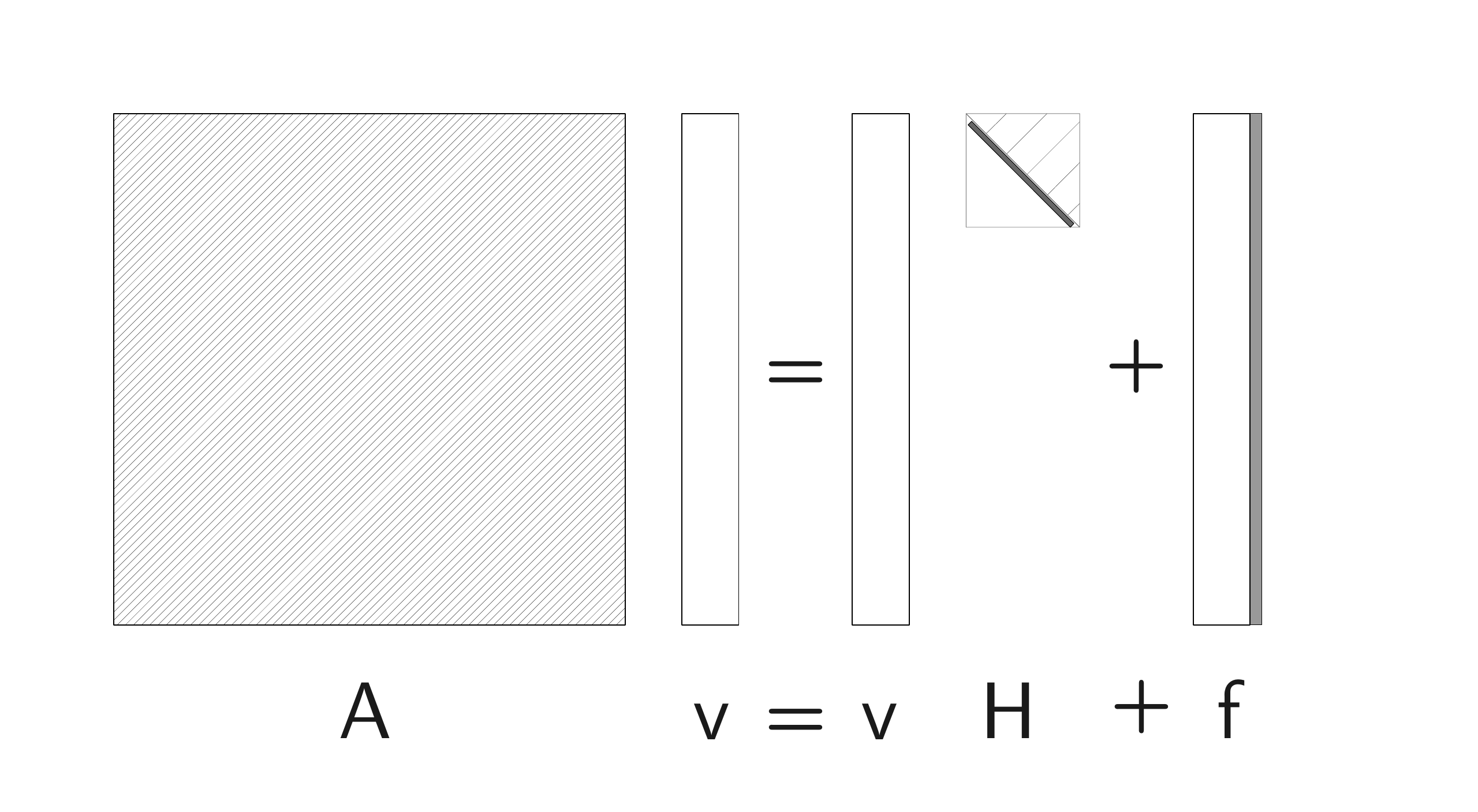}  
  \caption{Visualization of the Arnoldi relation, see \protect{\cite{Sorensen2002}}.}
  \label{fig:reduction_arnoldi}
\end{figure}

The basis $V_{m}$ of these subspaces can then be used to compute the Arnoldi relation \eqref{eq:arnoldi}, which projects a portion of $\mathcal{N}_{\mathrm{lin}}$ onto the Krylov subspace. This results in a Hessenberg matrix $H_{m}$ that is usually much smaller than $\mathcal{N}_{\mathrm{lin}}$ and contains an approximation of the eigenvalues and eigenvectors of the operator.
\begin{align}
 \mathcal{N}_{\mathrm{lin}} V_m = V_{m} H_{m} +  f_m e_m^{H}\label{eq:arnoldi}
\end{align}
In \eqref{eq:arnoldi}, $f_m e_m^{H}$ represents the last component of the residual vector of the size $m$ Arnoldi decomposition, which is visualized in Fig. \ref{fig:reduction_arnoldi}.

\subsection*{Implicitly restarted Arnoldi method}
In order to solve the eigenvalue problem \eqref{eq:stability} iteratively, the implicitly restarted Arnoldi method (IRAM) developed by \cite{Sorensen2002} is used. It offers a robust method for obtaining Ritz pairs - approximations to the Eigenvalues and Eigenvectors of the operator $\mathcal{N}$. Its main benefits are twofold. On one hand, it is a Krylov-type method and therefore does not need the explicit Jacobian of the operator. On the other hand, the IRAM allows a certain degree of control concerning which part of the spectrum is to be computed. This feature is especially important in order to obtain the desired results in a manageable time frame. In essence, it is a restarted Arnoldi method which allows the removal of unwanted Eigenvalue information during the restart of the iteration by using a QR-shift. A schematic overview of the algorithm is presented in algorithm \ref{alg:iram}. 
\begin{algorithm}
  \begin{algorithmic}
    \State Start with a $m = k + p$ step Arnoldi factorization (algorithm \ref{eq:arnoldi})
    \State $\mathcal{N}_{\mathrm{lin}} V_m = V_m H_m + f_m e_m^{*}$ \Comment{parallel}
    \While{not converged}
	\State compute $\mathrm{spectrum}(H_m)$, select $p$ shifts $\mu_1, \mu_2, \dots ,\mu_p$
	\State perform implicit QR-Shifts:
	\State $Q = I_m$
	\For{$j = 1,\dots,p$}
	\State   call qr factorization($H_m - \mu_j I$) $\rightarrow$ $[Q_j, R_j]$
	\State   $H_m  \gets Q_j^{*}H_m Q_j$
	\State   $Q_{  }  \gets Q Q_j$
	\EndFor
	\State update the Arnoldi factorization:
	\State$\hat{\beta}_k = H_m(k+1,k)$
	\State ${\sigma}_k = Q(m,k)$
	\State$f_k = v_{k+1} \hat{\beta}_k + f_m \sigma_k$
	\State $V_k \gets V_m Q(:,1:k)$
      \State  $H_k \gets H_m Q(1:k,1:k)$
      \State convergence check using \eqref{eq:ritz_estimate} \Comment{parallel}
	\State use obtained k-step factorization as starting point:
    \State $\mathcal{N}_{\mathrm{lin}} V_k = V_k H_k + f_k e_k^{*}$
    \State    apply p steps of Arnoldi method to obtain new factorization
    \State $\mathcal{N}_{\mathrm{lin}} V_m = V_m H_m + f_m e_m^{*}$ \Comment{parallel}
    \EndWhile
  \end{algorithmic}
   \caption{Implicitly restarted Arnoldi factorization \protect{\cite{Sorensen2002}}. Parallel parts are represented by a ``parallel'' comment. }  
  \label{alg:iram}
\end{algorithm}
An important aspect of the IRAM with respect to the stability analysis is to define a suitable convergence criterion to determine when the desired Ritz pairs have been computed. As pointed out by \cite{Lehoucq1996}, choosing an appropriate criterion is difficult for cases involving non-Hermitian operators which occur often when analyzing compressible flows since the residual norm $\|f_m\|$ of the Arnoldi decomposition does not necessarily imply convergence as opposed to the Hermitian case. To ensure a small backward error, a Ritz estimate \eqref{eq:ritz_estimate} proposed by \cite{Lehoucq1996} is used. This criterion is applied to each computed Ritz pair $(\hat{x}, \theta)$.
\begin{align}
 || f_m || \, |e_m^{T} \hat{x}| \leq \max(\epsilon_{\mathrm{m}} ||H_m||, tol_ {\mathrm{iram}} |\theta |) \label{eq:ritz_estimate}
\end{align}

\subsection*{Cayley transformation}
While the selective properties of the IRAM are very useful for this task, they are sometimes not sufficient when considering stability problems of compressible flows, which lead to problems due to clustering of Eigenvalues close to the origin or when dealing with badly conditioned flows. In order to enhance the selection properties of the framework even further, a Cayley transformation \eqref{eq:cayley_transform}, as proposed by \cite{Meerbergen1994}, is applied to the Eigenvalue problem \eqref{eq:stability}.
\begin{align}
 C(\mathcal{N}_{\mathrm{lin}}) = \left(\mathcal{N}_{\mathrm{lin}} - \sigma  I\right)^{-1} \left(\mathcal{N}_{\mathrm{lin}}-\mu I\right) \label{eq:cayley_transform}
\end{align}
This leads to a generalized Eigenvalue problem of the type
\begin{align}
 (\mathcal{N}_{\mathrm{lin}}-\sigma I)v_{j+1} = (\mathcal{N}_{\mathrm{lin}}-\mu I)v_j \text{, } \label{eq:cayley_arnoldi} 
\end{align}
which has to be solved.
The solution of this linear system requires an iterative solver. Due to the nature of the iterative solvers, the Cayley transformation is therefore only applied inexactly.

\begin{figure}
  \centering
  \includegraphics[width=0.9\textwidth]{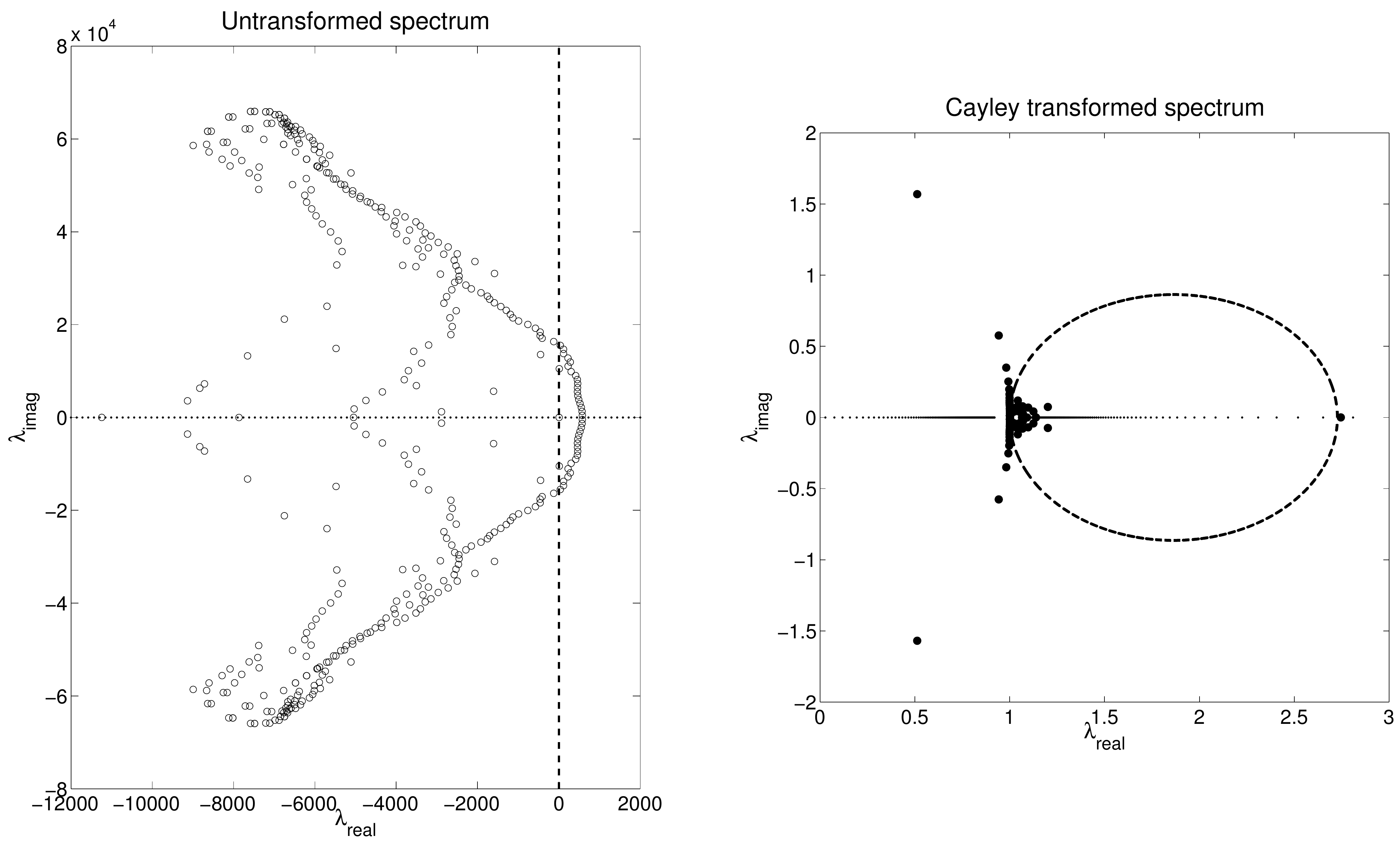}  
  \caption{Comparison between an untransformed spectrum and its Cayley transformed counterpart, using $\mu = 440.5 $, $\sigma = 1200.23 $. Also showing transformation of two Cartesian lines.}
  \label{fig:cayley_comparison}
\end{figure}

To illustrate how the Cayley transformation affects the spectrum of the transformed operator, Fig. \ref{fig:cayley_comparison} shows a comparison of a sample spectrum with and without a Cayley transformation. Depending on the parameters $\mu$ and $\sigma$, the transformation has two effects on the shape of the spectrum. It introduces a rotation as well as a stretching effect. Therefore, it can be used to separate clusters of Eigenvalues and to rotate desired Eigenvalues into the part of the spectrum with positive Eigenvalues, which are favored by Arnoldi based methods. Another important feature of this transform is to map inner Eigenvalues to the outer part of the spectrum. This is important since Arnoldi based method usually converge first onto Eigenvalues distant from the origin.

\subsection*{Iterative solvers}
The basic solver for the solution of \eqref{eq:cayley_arnoldi} is a typical restarted GMRES based on \cite{Saad1986}, which was chosen for its robust numerical behaviour. Other methods are also applicable, for example BiCGstab.

Since the solution of \eqref{eq:cayley_arnoldi} usually incurs a large numerical effort, it is beneficial to reduce the iterations required to solve it as much as possible. We investigate the possible speedup of a modified GMRES version and compared them to the standard formulation. The analyzed LGMRES uses error approximations to augment the Krylov subspace between restart cycles \cite{Baker2005}. The variant aims at improving the speed of convergence of the traditional restarted GMRES by lessening the impact of the necessary restart.

\subsection*{Parallelization}

In order to compute large scale stability analysis, parallelization of the framework is required. Since the framework is based on IRAM, it lends itself well to parallelization since the numerically most expensive steps, the construction of the new Krylov subspaces, can be distributed among available CPUs such that each process does not require information about the entire computational domain. This reduces the required communication between the processes significantly. The steps that require parallelization are marked in the respective algorithm scheme such as Listing \ref{alg:iram}. The implementation used in the scope of this work achieves parallelization by parallelizing the derivatives using MPI, but other means of parallelization are applicable as well.

\subsection*{Preconditioning}
Preconditioning techniques are often necessary to improve the speed of convergence when solving large linear systems arising during the stability analysis. While many methods for preconditioning are well researched, their application to a matrix free and parallel case is not straightforward. Since a global preconditioner is difficult to assemble due to the lack of the operator matrix, a local approach for the preconditioner was chosen for this work to serve as a proof of concept for matrix free preconditioning in parallel stability analysis. 

The framework uses a local block-Jacobi scheme similar to \cite{Hegland1991} which assembles a local preconditioner matrix on each process and uses it to accelerate the solution of the linear system. While this approach is rather simple, it serves as a starting point for future refinement.

\section*{Numerical framework overview}
\begin{figure}[htb]
  \centering
  \includegraphics[width=1\textwidth]{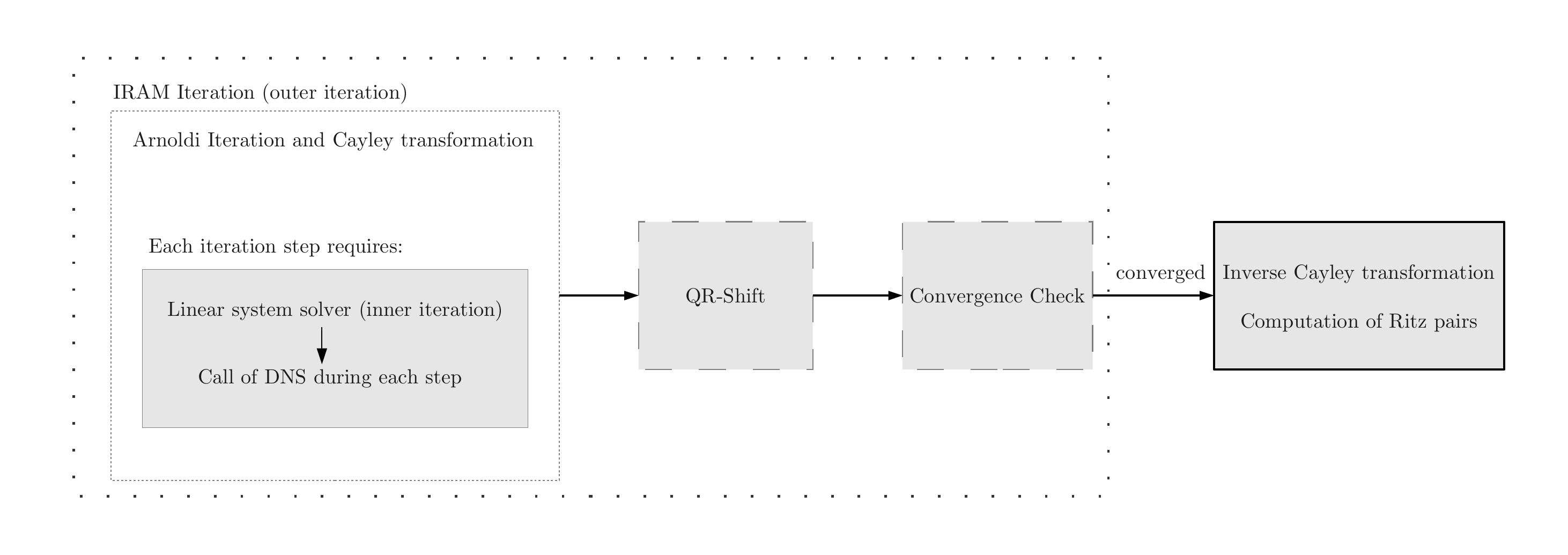}  
  \caption{Schematic overview of the stability framework.}
  \label{fig:flowchart_framework}
\end{figure}

The proposed framework can be roughly divided into two parts, an outer iteration loop consisting of the IRAM-algorithm \cite{Sorensen2002}, which obtains the stability information, and an inner iteration loop for the solution of the linear system required for the application of the inexact Cayley transformation. The relationship between these is depicted in Fig. \ref{fig:flowchart_framework}. \par
An application of the operator $\mathcal{N}_{\mathrm{lin}}$ is required for each step of the inner iteration. Each of these calls of the operator translates to one evaluation of the defining equations \eqref{eq:navier_stokes}. This is usually the step which requires the most computational resources. It is also the ``innermost'' step of the framework and therefore the most called.

\section{Model problem}
\begin{figure}[htb]
  \centering
  \includegraphics[width=0.55\textwidth]{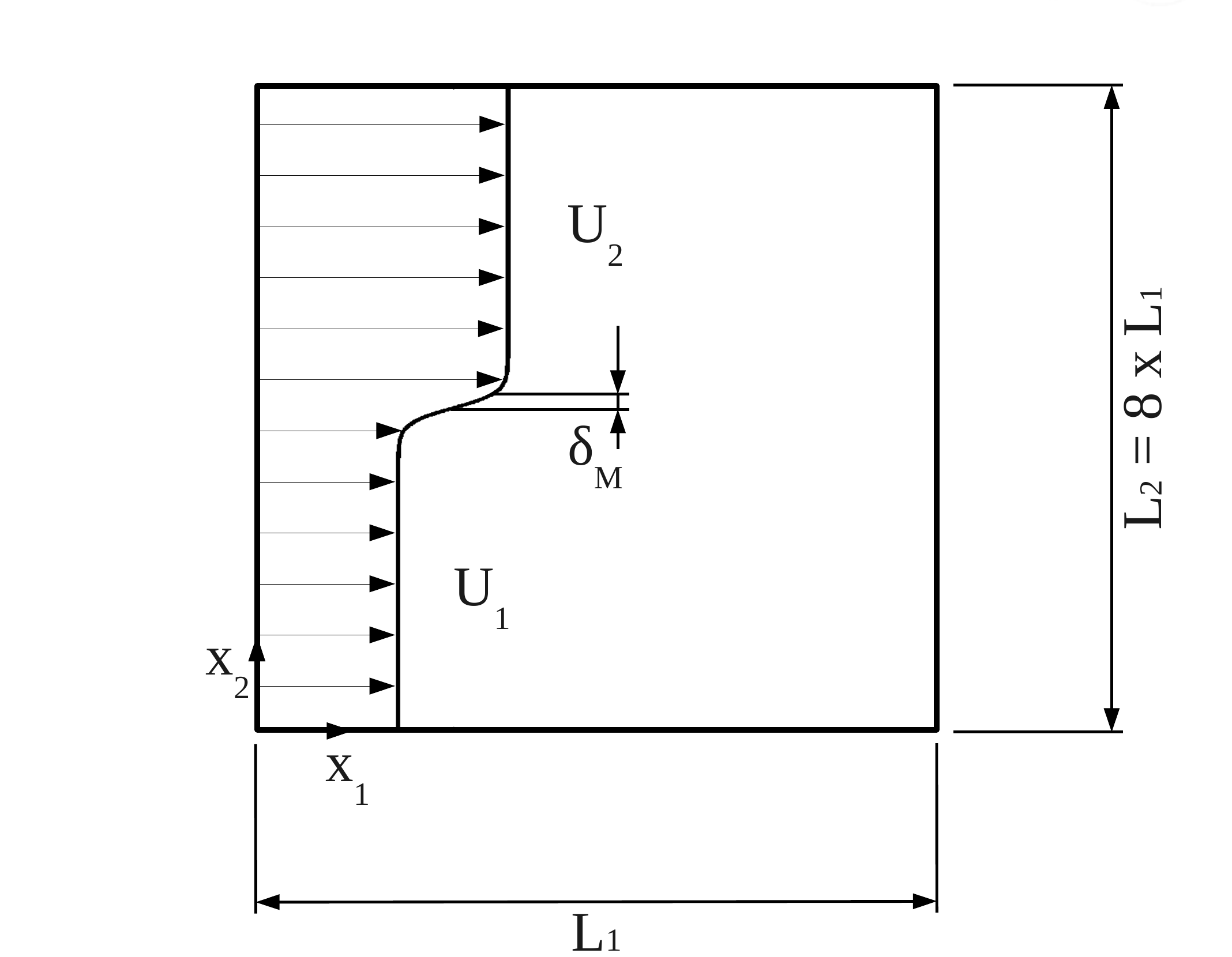}  
  \caption{Draft of the used computational domain showing dimensions and the hyperbolic tangent profile.}
  \label{fig:domain}
\end{figure}

To illustrate the proposed framework, a model problem is used in the scope of this paper. We investigate the temporal stability of two-dimensional compressible mixing layer using a hyperbolic tangent velocity profile similar to the works of Blumen \cite{Blumen1970}. The profile used is described by equation \eqref{eq:tanh}, which imposes a velocity profile in the $x_{1}$-component of the velocity onto the stream. The thickness of the shear layer is controlled by the parameter $\delta$.
\begin{align}
  u_1(x_{2}) = u_{\infty} \tanh\left(\frac{x_{2} - 0.5 \cdot L_{2}}{\delta}\right) \label{eq:tanh},
\end{align}
with $u_\infty$ as reference velocity and $L_2$ as domain length in $x_2$ direction.
The model problem is set up in a computational domain with $128\times512$ points as depicted in Fig. \ref{fig:domain}. It employs a periodic boundary condition on the left and right boundaries as well as a non-reflective boundary condition on the lower and upper boundaries. Additionally, grid stretching was used to improve the resolution in the critical parts of the domain where the mixing effect occurs. This leads to a minimum cell distance in $x_{2}$ - direction of $\approx 0.08 \delta$ centered towards to highest velocity gradients present in the profile.

The tangent hyperbolic velocity profile is seeded into the domain as a starting condition. In order to test the parallel aspect of the framework, the computation is distributed between at least 4 processors. 

This model problem is investigated with regard to its linear temporal stability. The goal is to identify the least stable mode by using the previously outlined framework.
The domain length in $x_{1}$-direction $L_{1}$ should fit at least one entire wavelength of the desired mode, and was chosen to be exactly one wavelength of the least stable mode scaled by mixing layer thickness $\delta$. Assuming a wavenumber $\alpha$ of the least stable mode, the length of the domain in $x_{1}$-direction would result as $L_{1}\frac{2 \pi \delta}{\alpha}$. The domain length in $y$-direction was chosen as $8$ times $L_1$.

Compared to the traditional matrix based approach on stability analysis, the presented matrix free framework requires significantly less memory. Consider the above model problem with a $128\times512$ points resolution. Assuming double precision, storing the operator alone would require $(128\cdot512)^{2} \cdot 64 \mathrm{Bit} \approx 32 \mathrm{GB}$ of memory. When performing a parallel computation, this requirement is shared between all available nodes. The memory requirements of the matrix-free approach are tested in the following Sec. \ref{sec:results_memory}.

%% file: 04_results.tex
\section{Validation}
The proposed framework is validated using the model problem described in the previous section. The least stable modes for a various settings are computed and their growth rates compared to literature, see \cite{Blumen1970,Mack2009}. All computations are performed in parallel and cover the different types of iterative solvers described in the previous section as well as the preconditioner introduced previously.

\subsection*{Results - Validation}
\begin{figure}
  \centering
  \includegraphics[width=1\textwidth]{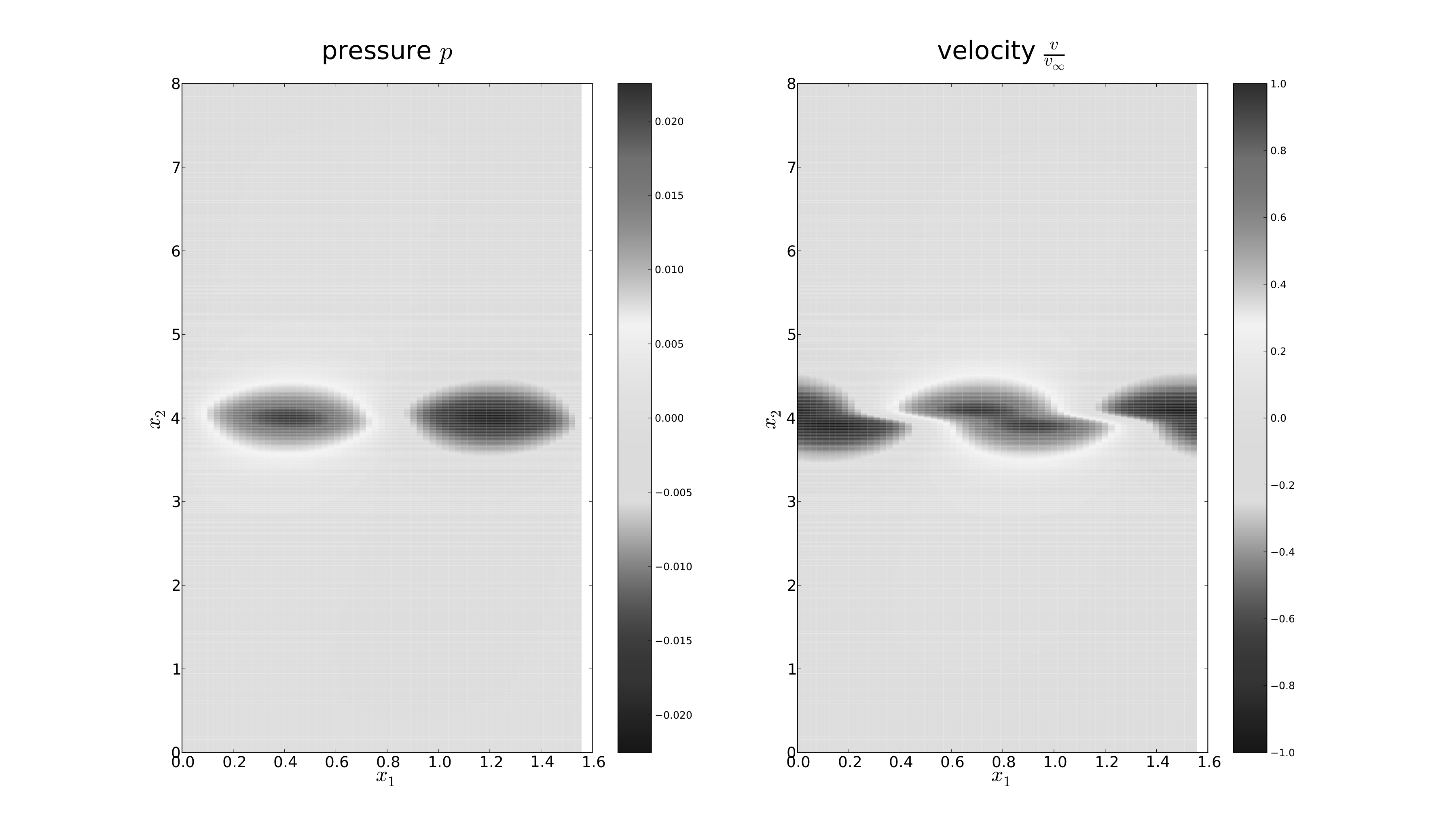}
  \caption{Least stable mode computed with resolution $128\times512$ using 64 CPUs. One period in x-direction is shown. Depicted are fields pressure $p$ and the velocity component $u_2 = v$. Also, the figure shows the underlying grid used in the computation.}
  \label{fig:mode}
\end{figure}

The proposed framework is able to compute growth rates of the least stable mode in good accordance to the results given by Blumen \cite{Blumen1970} and Mack \cite{Mack2009}. The detailed results are compared in Tbl. \ref{tab:validation_results}. The different configurations are designed to cover a range of Mach numbers as well as different resolutions at a high Reynolds number. The different configuration parameters are listed in Tbl. \ref{tab:validation_config}.
In addition to the computed growth rates, an exemplary mode computed during the validation is presented in Fig. \ref{fig:mode}.

From this results we find the growth to be computed in good accordance with literature. 
\begin{table}[htb]
  \begin{center}
  \begin{tabular}{cccc}

    Configuration & $M$ & $\mathrm{Re}$ & Resolution \\   \hline
    V01 & 0.1 & $ \rightarrow \infty$ & $64\times 256$  \\
    V02 & 0.5 & $ \rightarrow \infty$ & $64\times 256$  \\
    V03 & 0.9 & $ \rightarrow \infty$ & $64\times 256$  \\
    V04 & 0.5 & $ 1000 $ & $64\times 256$  \\
    
    V05 & 0.1 & $ \rightarrow \infty$ & $16\times 64$ \\
    V06 & 0.5 & $ \rightarrow \infty$ & $16\times 64$ \\
    V07 & 0.9 & $ \rightarrow \infty$ & $16\times 64$  \\
    V08 & 0.5 & $ 1000 $ & $16\times 64$  \\        
  \end{tabular} 
  \caption{Configuration parameters for different validation setups.}
  \label{tab:validation_config}
  \end{center}
\end{table} 
\begin{table}[htb]
  \begin{center}
  \begin{tabular}{cccc}
    Configuration & computed growth rate $\omega$ & $\omega$ by Blumen & $\omega$ by Mack  \\   \hline
    V01 & $0.189$ & $0.187$ & $0.187521$ \\
    V02 & $0.1411$ & $0.141$ & $0.1411$ \\
    V03 & $0.0547$ & $0.055$ & $0.054723$ \\
    V04 & $0.1273$ & n/a & $0.1271$  \\%
    
    V05 & $0.189$ & $0.187$ & $0.187521$  \\
    V06 & $0.139$ & $0.141$ & $0.141167$\\
    V07 & $0.0540$ & $0.055$ & $0.054723$ \\
    V08 & $0.1270$ & n/a & $0.1271$  \\
  \end{tabular} 
  \caption{Results of the validation for various setups.}
  \label{tab:validation_results}
  \end{center}
\end{table} 

\section{Numerical Experiments}
As mentioned previously, the computational costs and required memory of the stability problem increases rapidly when considering problems which require high resolutions. While the memory requirements are mitigated to a great extent by the use of a matrix-free framework, the computational effort required is still significant. It is therefore important to reduce this computational effort, for example by improving the convergence behavior of the stability framework to avoid unnecessary iterations.

This section therefore investigates the effectiveness of the preconditioning method and iterative solvers with regard to speed up the iteration based on the results of numerical experiments. 
Since one of the main objectives of the presented framework is its matrix-free and parallel implementation, we will discuss the scalability and especially memory aspects based on the numerical experiments.

Lastly, while the choice of Cayley parameters has an influence on the performance of the framework, the presented numerical experiments do not cover a study on these parameters, since the focus of this work is to provide an overview of the framework. For a more detailed description of the effects of the Cayley transform, the reader is referred to \cite{Meerbergen1994} or \cite{Mack2009}. Still, a few general remarks on the choice of Cayley parameters for this case are summarized at the end of this section.

The numerical experiments are based on the model problem discussed previously and differ in resolution, type of preconditioning and iterative solver used to cover a variety of possible combinations.

All further experiments share the following basic parameters: A Mach number $\mathrm{M} = 0.5 $, Reynolds number $\mathrm{Re} =  \rightarrow \infty$, domain length in $x_{1}$- direction $L_{1} = 15.83 \delta$, tolerance of the IRAM $tol_{\mathrm{iram} = 1e-5}$  and tolerance for the GMRES-type method  $tol_{\mathrm{gmres}}= 1e-6$. The subspace size $m$ of the IRAM was varied from $100 - 250$, depending on the resolution of the case, with lower resolutions using smaller subspace sizes.

\subsection*{Preconditioning Scheme}
Aside from the growth rate as a measure of the correctness of the computation, the number of required matrix-vector product operations is recorded. Since these correlate directly with the amount of required operator evaluations - the by far most numerically expensive part in these simulations. They serve as a good measure to compare different methods regarding speedup.
\begin{figure}
  \centering
  \includegraphics[width=\textwidth]{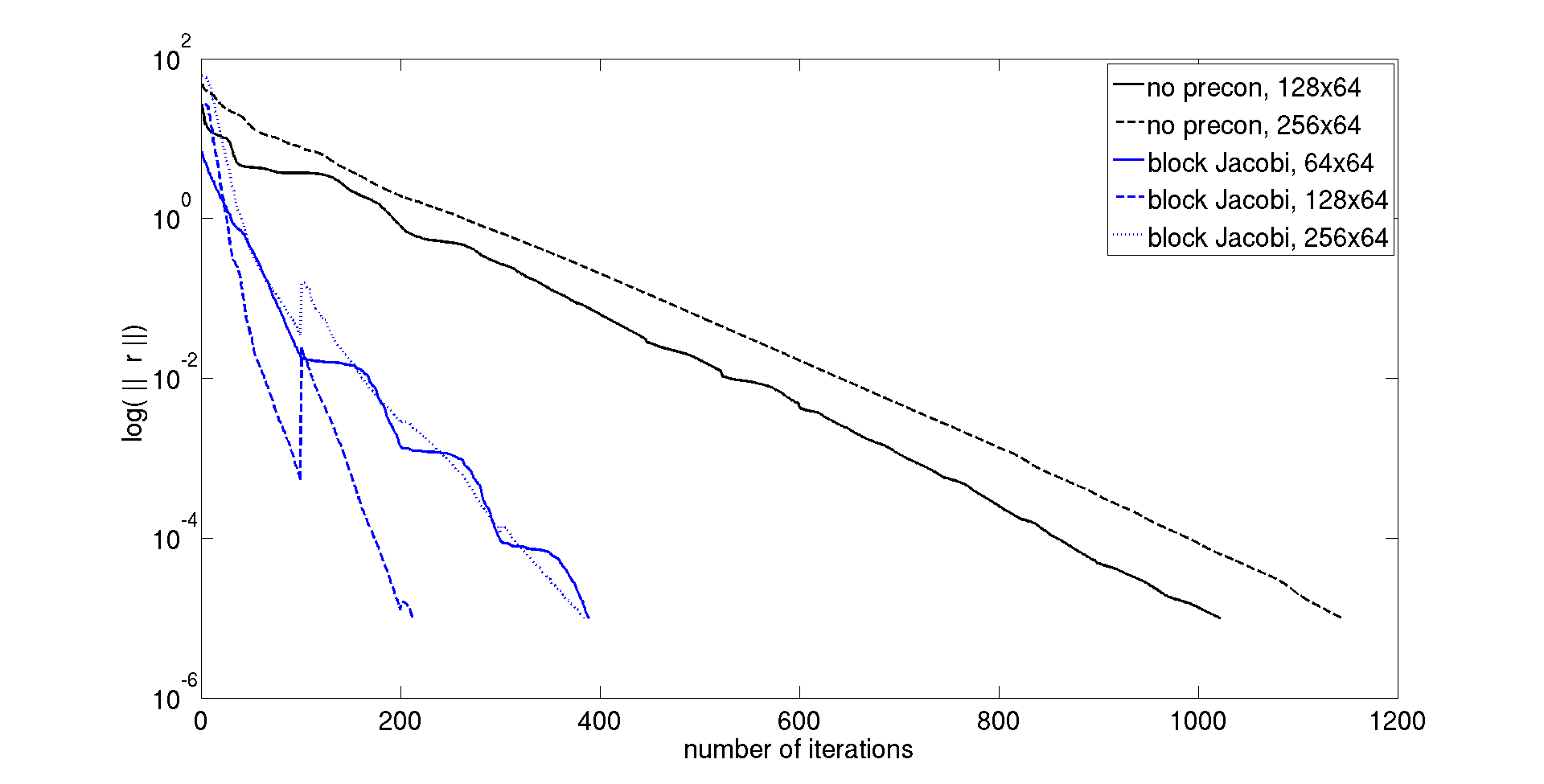}
 \caption[Comparison of convergence with preconditioning vs. no preconditioning]{Comparison of the iteration process of one call of the linear solver for five different setups using LGMRES.}
 \label{fig:precon_comparison}
\end{figure} 
Fig. \ref{fig:precon_comparison} compares different computations,  using the Block Jacobi preconditioner as well as running without preconditioning. A reduction in needed iterations for the preconditioned cases is clearly visible in the comparison. The reduction differs from case to case, but usually ranges 40-60\%. 

Aside from the speedup aspect another equally important aspect of the preconditioning became apparent during the numerical experiments. Since the stability problem arising from compressible flows is usually conditioned very poorly, computations without preconditioning often fail to produce accurate results. The use of a preconditioner alleviated this problem by enabling the computation of viable results in these cases and making the computation more robust. The resolutions used for the results above was therefore chosen small to allow a comparison to the unpreconditioned case. 

\subsection*{Linear Solvers}
Aside from the speedup achieved by the preconditioning, the speedup of the LGMRES solver is also investigated by numerical experiments. Table \ref{tab:comparison_overall} has exemplary results for the effectiveness of the LGRMES-method compared to standard GMRES. 

In comparison to the effectiveness of the preconditioner, the speed up achieved by this method is rather small. Given the very low additional numerical effort incurred by this method compared to standard GMRES, the use of LGMRES is still advisable.

\subsection*{Memory and computation time considerations} \label{sec:results_memory}
In order to highlight the effects of the different methods with respect to speedup, a comparison between selected setups is displayed in Tbl. \ref{tab:comparison_overall}. The table compares the different configurations in terms of required iterations as well as required memory during the computation. Note that the computation using the higher resolution was not possible without the use of a preconditioner. The memory requirement displayed therein is estimated analytically.
\begin{table}[htb]
  \begin{center}
    \begin{tabular}{lccccc}
			& resolution 		& GMRES 	& memory$^*$ 			& LGMRES 	& memory$^*$ 			\\ \hline
    no precond. 	& $256\times 64$  	&  1419325  	&  $ 0,1 \mathrm{GB}^{*}$  	& 1389364 	&  $ 0,1 \mathrm{GB}^{*}$  	\\ 
    no precond. 	& $1024\times 128$  	& / 		&  				&/  		& 				\\
    block Jacobi 	& $256\times 64$        & 65164 	& $ 4.1 \mathrm{GB}$ 		& 63472 	&  $ 4.1 \mathrm{GB}$ 		\\ 
    block Jacobi 	& $1024\times 128$      & 1459950  	& $ 256,4 \mathrm{GB}$ 		& 1411357 	& $ 256,4 \mathrm{GB}$ 
  \end{tabular} 
  \caption{Iteration times for different solvers, preconditioning methods and resolutions for the model problem. $(^*)$ The memory requirement is shared across the nodes.}
  \label{tab:comparison_overall}
  \end{center}
\end{table} 
The table contains several important aspects of the framework. The first to be noted is the memory requirement of the method is heavily influenced by the method of preconditioning. In its current iteration, the use of the Block Jacobi method requires a significant amount of memory when investigating higher resolutions compared to the unpreconditioned case. This is a trade-off for the drastic reduction in required iterations compared to the unpreconditioned computations. For the examples provided in the given table, the preconditioned computation required roughly $20$ times less iterations, a significant speedup even when considering the memory requirements of this method. However, the memory requirement is shared over the parallelized nodes.

\begin{figure}
  \centering
  \includegraphics[width=0.55\textwidth]{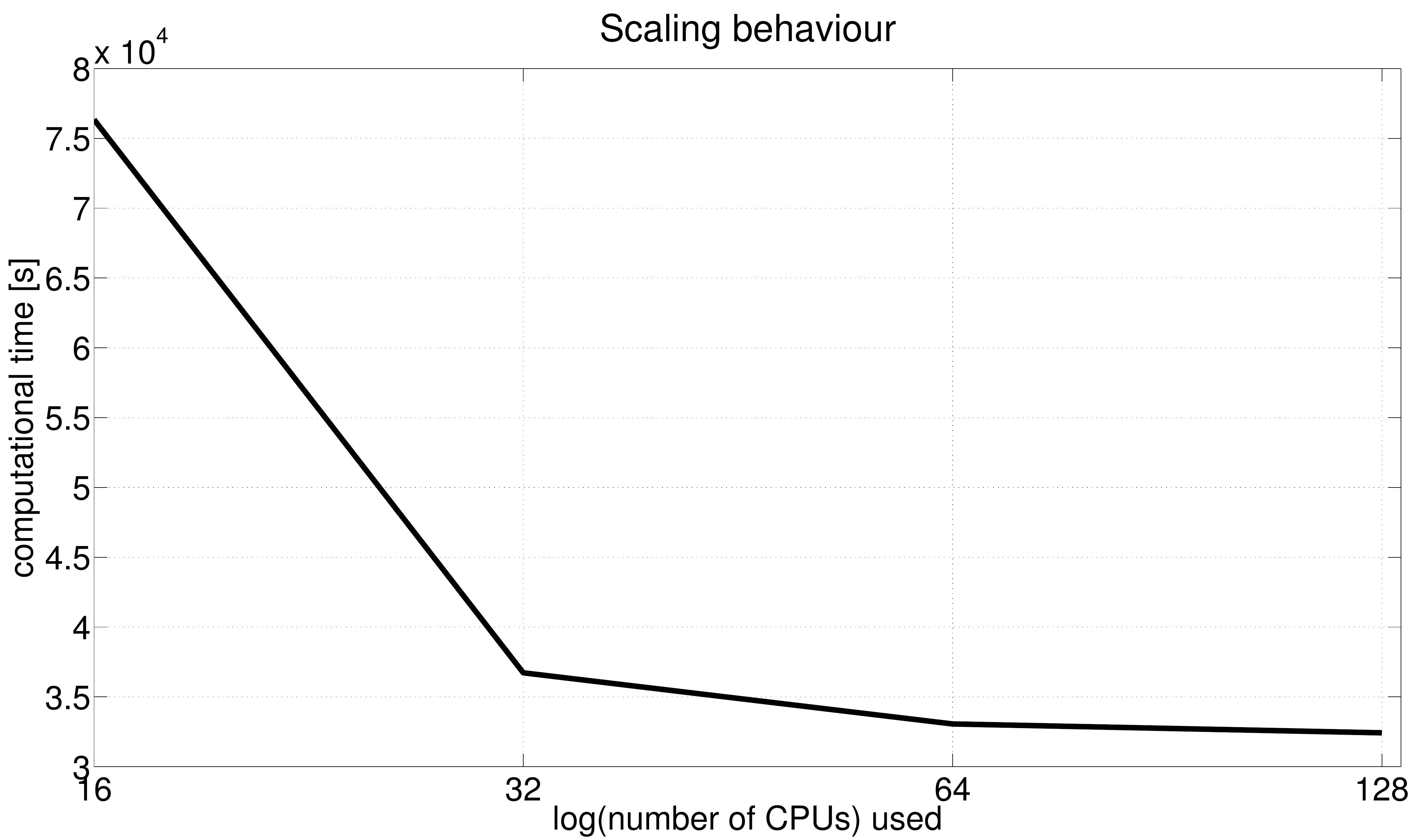}
  \caption{Scaling behavior of the framework.}
  \label{fig:scaling}
\end{figure}

As an additional metric, the scaling of the computational time and the memory requirement is collected in Fig. \ref{fig:scaling} to give an example for the scaling behavior of the framework when using the Block-Jacobi preconditioner. 
The scaling observed in this case massively deteriorates from $64$ processors onward. This can be explained by the implementation of the preconditioner which requires a huge amount of calls of $\mathcal{N}(q)$ (ergo the right hand side of the considered equation) to compose the preconditioner prior to its application. With increasing number of processors used, the numerical costs of the composition surpass the costs of the solution of the preconditioned linear system, since the calls of $\mathcal{N}(q)$ require MPI calls.  While this process is parallelized, it still acts as a bottleneck when using a relatively large number of processors - for the parameters given, this number is at $64$ processors. Furthermore, the preconditioner cannot be omitted despite deteriorating the scaling of the framework since the preconditioner is a requirement to enable the correct computation in the first place. 

Due to this behavior, the improvement of the preconditioner seems to be the most promising route to consider for improving the performance of the framework. As with the framework in general, the goal is to reduce the number of required right hand side calls. This could be achieved by a using non compact, lower order discretization of the operator $\mathcal{N}(q)$. This would significantly reduce the MPI communication during operator calls while also having the possible benefit of reducing the memory required during the composition of the preconditioner, \cite{Mack2009}.
\begin{figure}
  \centering
  \includegraphics[width=0.55\textwidth]{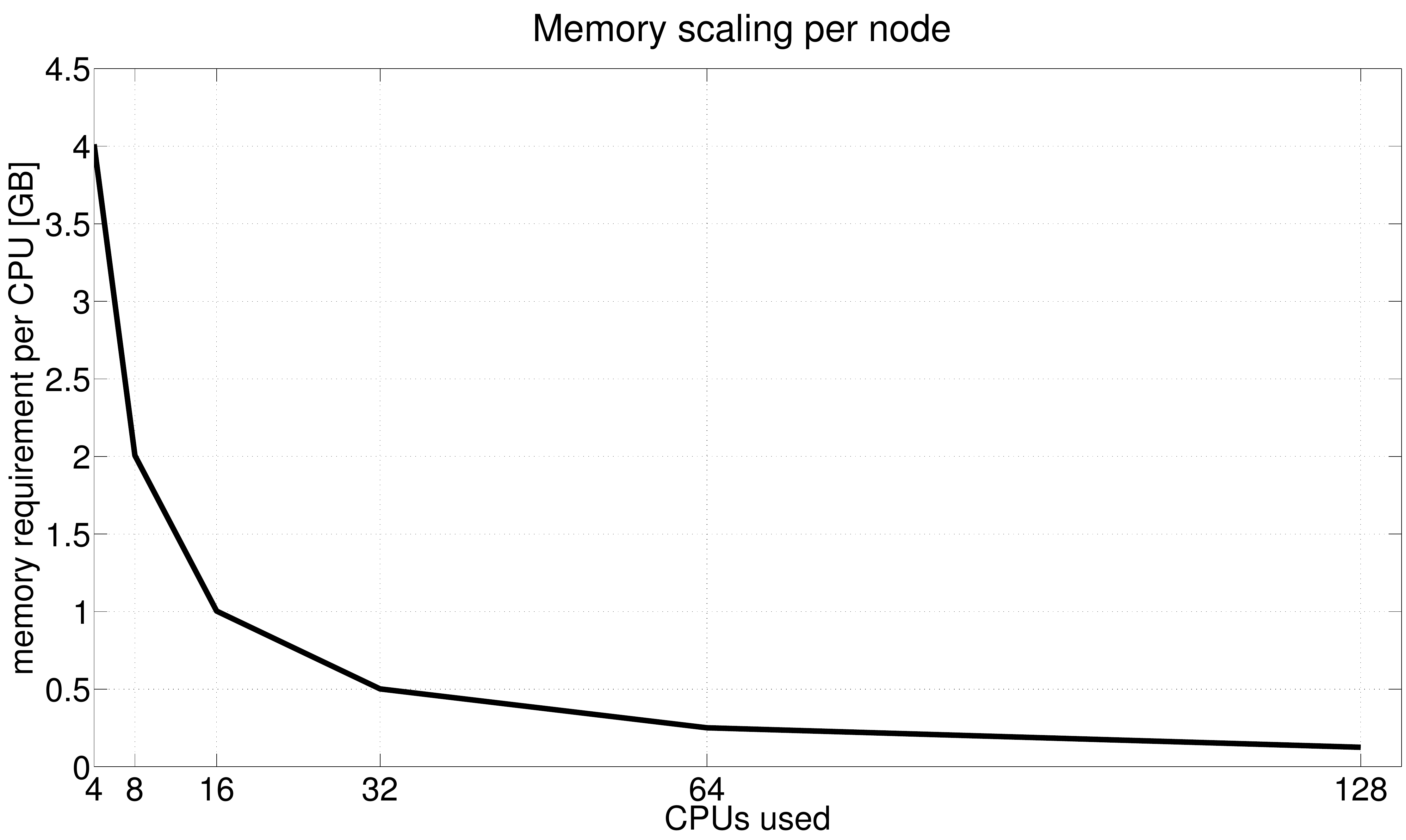}
  \caption{Theoretical memory scaling of the framework with preconditioning.}
  \label{fig:memory}
\end{figure} 
Fig. \ref{fig:memory} shows the estimated memory requirement per node when using preconditioning. 
The actual memory requirements during numerical experiments the  scale nearly linearly with the number of used processors, except for a small overhead which results in a sub-linear scaling. This is likely dependant on the method of parallelization used in the framework, although this was not investigated in the scope of this work.

\subsection*{Cayley parameters and further remarks}
The numerical experiments performed for this work generated further observations which are not covered in the above sections, yet can still be useful to get a better understanding of the problem at hand and to serve as guidelines for common problems.

Firstly, the condition of the linear systems deteriorates with increasing resolution. This is to be expected since an increase in resolution leads to a higher density of computable modes in the mixing layer. This results in an increased clustering of Eigenvalues which can affect the condition of the linear system. Additionally, a higher resolution can also result in an increase of unstable modes introduced by the discretization scheme. There are several ways to lessen the impact of these phenomena.

A first measure to be considered when dealing with the increased resolution is to increase the sizes of the subspaces available to the IRAM and GMRES. This step is usually required to account for the increased complexity of the computed spectrum and comes with an increase in memory and computational resources. For the resolutions investigated in this works, subspace sizes varying from $100$ to $250$ were found adequate. 

In addition to the increase in subspace size, an adjustment of the Cayley parameters can also lead to an improvement of the computation for higher resolutions. Unfortunately, the method for determining the correct parameters is not as straightforward compared to the subspace size since the Cayley parameters are seemingly not correlating with the resolution in a simple fashion. For the resolutions considered herein, the following parameter sets in Tbl. \ref{tab:cayley_parameters} have been used successfully.
\begin{table}[htb]
  \begin{center}
  \begin{tabular}{cc}
    resolution & recommended parameters \\ \hline
    $32 \times 128$ &$\mu = 230.5 $, $\sigma = 800.32 $ \\
    $32 \times 256$ &$\mu = 444.5 $, $\sigma = 1200.32 $ \\
    $64 \times 512$ &$\mu = 444.5 $, $\sigma = 1200.32 $ \\
    $128 \times 1024$ &$\mu = 444.5 $, $\sigma = 1200.32 $
  \end{tabular} 
  \caption{Recommended parameters for different resolutions of the mixing layer model problem. }
  \label{tab:cayley_parameters}
  \end{center}
\end{table} 
While the Cayley parameters can influence the number of iterations required by the linear solver, the effect of fine tuning the parameters was not found significant.

\section{Conclusions}
A framework for the matrix free and parallel stability analysis of compressible flows has been presented and validation results and speedup experiments have been showcased. To enhance the performance of the framework, a straightforward parallel preconditioning method was presented and tested successfully. The methods presented herein can be seen as a proof of concept for the matrix free stability analysis in parallel environments.

Furthermore, the numerical experiments provide several ways that can be explored for a further increase in efficiency of the computation. As mentioned previously and judging from the results of the numerical experiments, the preconditioning scheme appears to be the most promising route to explore for further improvements. Two aspects can be considered as an angle for improvements to the preconditioning scheme.

Firstly, the construction of the preconditioner itself could be handled in a more sophisticated manner regarding memory requirements and method of construction. For example, rather than using the whole local operator as base for the preconditioning matrix, one could revert to using a reduced operator which still covers the physics but uses a much lower order finite difference scheme, as employed by \cite{Mack2009} in the non parallel case. This would reduce the memory required by the preconditioner while also facilitating its construction.

Secondly, the inversion and application of the preconditioning matrix offers room for improvement, e.g. by using sparse techniques for saving the matrix if the structure of the matrix is suitable. 

While two different linear solvers were compared, the impact of the chosen method was found to be small compared to the impact of the preconditioning method. Therefore, while other methods could still be investigated with respect to speed up, the study of the preconditioning method seems to be more efficient, especially.

\subsection*{Acknowledgements}
The authors gratefully acknowledge support by the
Deutsche Forschungsgemeinschaft (DFG) as part of
collaborative research center SFB 1029  ``substantial
efficiency increase in gas turbines through direct use
of coupled unsteady combustion and flow dynamics''